\documentclass{article}
\usepackage{spconf,amsmath,graphicx}
\usepackage{amssymb}
\usepackage{gensymb}
\usepackage{color}
\usepackage{booktabs}
\usepackage{subcaption}
\usepackage{adjustbox}
\usepackage{algorithm}
\usepackage{algorithmicx}
\usepackage{algpseudocode}
\usepackage{multicol}   
\usepackage{multirow}   
\usepackage{url}
\usepackage{hyperref}

\def\0{{\mathbf 0}}
\def\1{{\mathbf 1}}

\def\l{{\mathbf l}}

\def\v{{\mathbf v}}
\def\x{{\mathbf x}}

\def\y{{\mathbf y}}

\def\A{{\mathbf A}}

\def\D{{\mathbf D}}

\def\H{{\mathbf H}}
\def\I{{\mathbf I}}

\def\L{{\mathbf L}}

\def\cN{{\mathcal N}}

\def\ie{{\textit{i.e.}}}

\title{Joint Demosaicking / Rectification of Fisheye Camera Images using Multi-color Graph Laplacian Regularization}
\name{Fengbo Lan$^{\star}$, Cheng Yang$^{\star}$, Gene Cheung$^{\star}$, Jack Z. G. Tan$^{\dagger}$}
\address{$^{\star}$Dept of EECS, York University, Toronto, Canada \\ 
$^{\dagger}$Kandao Technology, Sydney, Australia \thanks{This research is partially funded by Kandao Technology.}}
\begin{document}
\ninept
\maketitle
\begin{abstract}
%\vspace{-0.05in}
%Fisheye cameras are essential for a wide range of virtual reality (VR) applications. 
To compose a 360\degree image from a rig with multiple fisheye cameras, a conventional processing pipeline first performs demosaicking on each fisheye camera's Bayer-patterned grid, then translates demosaicked pixels from the camera grid to a rectified image grid---thus performing two image interpolation steps in sequence. 
Hence interpolation errors can accumulate, and acquisition noise in the captured pixels can pollute neighbors in two consecutive processing stages.
In this paper, we propose a joint processing framework that performs demosaicking and grid-to-grid mapping simultaneously---thus limiting noise pollution to one interpolation. 
Specifically, we first obtain a reverse mapping function from a regular on-grid location in the rectified image to an irregular off-grid location in the camera's Bayer-patterned image.
For each pair of adjacent pixels in the rectified grid, we estimate its gradient using the pair's neighboring pixel gradients in three colors in the Bayer-patterned grid.
We construct a similarity graph based on the estimated gradients, and interpolate pixels in the rectified grid directly via graph Laplacian regularization (GLR).
Experiments show that our joint method outperforms several competing local methods that execute demosaicking and rectification in sequence, by up to 0.52 dB in PSNR and 0.086 in SSIM on the publicly available dataset, and by up to 5.53dB in PSNR and 0.411 in SSIM on the in-house constructed dataset.
\end{abstract}
\begin{keywords}
Fisheye camera, demosaicking, image rectification, graph signal processing. 
\end{keywords}
%
%\vspace{-0.1in}
\section{Introduction}
\label{sec:intro}
%\vspace{-0.05in}
Known for their ultra-wide fields of view, fisheye cameras play an important role in modern virtual reality (VR) applications \cite{8337839}.
In particular, when constructing a 360\degree\, image using a rig with multiple fisheye cameras capturing different viewpoint images \cite{8337839}, a conventional processing pipeline first performs demosaicking \cite{784434} on each camera's captured Bayer-patterned image (\ie, interpolate missing color components at each pixel location), then translates demosaicked pixels from the fisheye image grid to a target rectified image grid for consumption on a head-mounted display.
See \cite{8337839} for a detailed description of the image processing pipeline.

Performing demosaicking and rectification in sequence means that image interpolation is executed \textit{twice}, where at each stage, errors can accumulate, and acquisition noise at the captured pixels can smear neighbors, resulting in correlated noise and degrading output image quality.
To alleviate this problem, in this paper we propose a joint processing framework that performs demosaicking and grid-to-grid mapping simultaneously for fisheye camera images.
This means that image interpolation is performed only \textit{once}, limiting the effect of noise pollution during only one interpolation step.
\textit{To the best of our knowledge, we are the first to perform joint demosaicking / rectification in the fisheye camera image processing literature.}

Our work shares a similar optimization philosophy with recent efforts on joint demosaicking / denoising (JDD) \cite{1658081,4287011,Gharbi:2016:DJD:2980179.2982399,DBLP:journals/corr/abs-1802-04723}, that seek advantages over traditional separate demosaicking / denoising approaches, by addressing acquisition noise directly during the demosaick process that obscures noise.
Our problem is more complicated in that the mapping from the fisheye camera image grid to the rectified image grid typically involves non-integer locations. 
We address this issue with a graph-based approach that can more flexibly handle irregular sample placements.

Specifically, we first obtain a reverse mapping function from an integer pixel location in the rectified grid to a non-integer, irregular location in the camera's Bayer-patterned grid. 
For each adjacent pixel pair in the rectified grid, we estimate its gradient using the pair's neighboring captured pixel gradients in the Bayer-patterned grid in all three colors, accounting for local inter-color correlations.
We construct a similarity graph based on the estimated gradients, and interpolate pixels in the rectified grid directly via \textit{graph Laplacian regularization} (GLR) \cite{7814302}.
Experiments show that our joint method outperforms several competing local methods that execute demosaicking and rectification in sequence, 
by up to 0.52 dB in PSNR and 0.086 in SSIM on the Multi-FoV dataset \cite{7487210} and by up to 5.53dB in PSNR and 0.411 in SSIM on the in-house constructed dataset.
% 0.52 dB in PSNR and 0.08 in SSIM, respectively.

\vspace{0.05in}
\noindent
\textbf{Related Work}:
There are recent efforts to address inverse imaging problems using graph spectral techniques, including image denoising \cite{7814302}, soft decoding of JPEG images \cite{7740964, hu2015graph}, bit-depth enhancement \cite{wan2016image}, image deblurring \cite{8488519} and join dequantization / contrast enhancement \cite{8476571}.
Though the details of the employed graph-based regularizers differ across applications, the key to good reconstructed image quality remains the same: to appropriately design an underlying graph that captures inter-pixel similarities of the target image patch. 
Our current work differs from these works in that we estimate the gradient (difference) of a target color pixel pair in the rectified image grid using relevant observed gradients in the Bayer-patterned grid in all three color components.

%The outline of the paper is as follows. We first overview related works in Section\;\ref{sec:related}. We discuss the mapping function from pixels in the Bayer-patterned image sensor to the target rectified image grid in Section\;\ref{sec:overview}. We formulate and solve the image restoration problem in Section\;\ref{sec:formulate}. Finally, experimental results and conclusion are presented in Section\;\ref{sec:results} and \ref{sec:conclude}, respctively. 

%\vspace{-0.1in}
%\section{Related Work}
%\label{sec:related}
%\input{related.tex}

%\section{System Overview}
%\label{sec:overview}
%\input{overview.tex}

\vspace{-0.1in}
\section{Preliminaries}
\label{sec:prelim}
%\vspace{-0.05in}
We define basic definitions in \textit{graph signal processing} (GSP) \cite{ortega18ieee} to facilitate understanding in the sequel.
A graph $\mathcal{G}=(\mathcal{V}, \mathcal{E})$ consists of a node set $\mathcal{V}$ of size $N$ and an edge set $\mathcal{E}$ specified by $(i,j,w_{ij})$, where $i,j \in \mathcal{V}$ and $w_{ij} \in \mathbb{R}^{+}$ is a positive weight of an edge $(i,j)$ reflecting the similarity between samples at nodes $i$ and $j$. 
We define a symmetric \textit{adjacency matrix} $\mathbf{W} \in \mathbb{R}^{N \times N}$, where $W_{ij} = w_{ij}$ if $(i,j) \in \mathcal{E}$, and $W_{ij} = 0$ otherwise. 
We next define a diagonal \textit{degree matrix} $\mathbf{D} \in \mathbb{R}^{N \times N}$, where $\mathbf{D}_{ii}=\sum_{j} w_{ij}$. 
Given $\mathbf{W}$ and $\mathbf{D}$, we define a \textit{combinatorial graph Laplacian matrix} as $\mathbf{L}=\mathbf{D}-\mathbf{W}$.
By assigning sample value $x_i \in \mathbb{R}$ to node $i$, the ensemble $\x = [x_{1} \hdots x_{N}]^{\top} \in \mathbb{R}^N$ is called a \textit{graph signal} on graph $\mathcal{G}$.  
$\x^\top\L \x =\sum_{(i,j) \in \mathcal{E}} w_{ij}(x_i-x_j)^2$ is known commonly as the \textit{graph Laplacian regularizer} (GLR)~\cite{7814302} and is one smoothness measure of signal $\x$ with respect to the underlying graph defined by $\mathbf{L}$.

\vspace{-0.1in}
\section{Problem Formulation}
\label{sec:formulate}
%\vspace{-0.05in}
\subsection{Reverse Mapping of Output-Input Image Grids}

\begin{figure}[!ht]
	\centering
	\includegraphics[width=7cm]{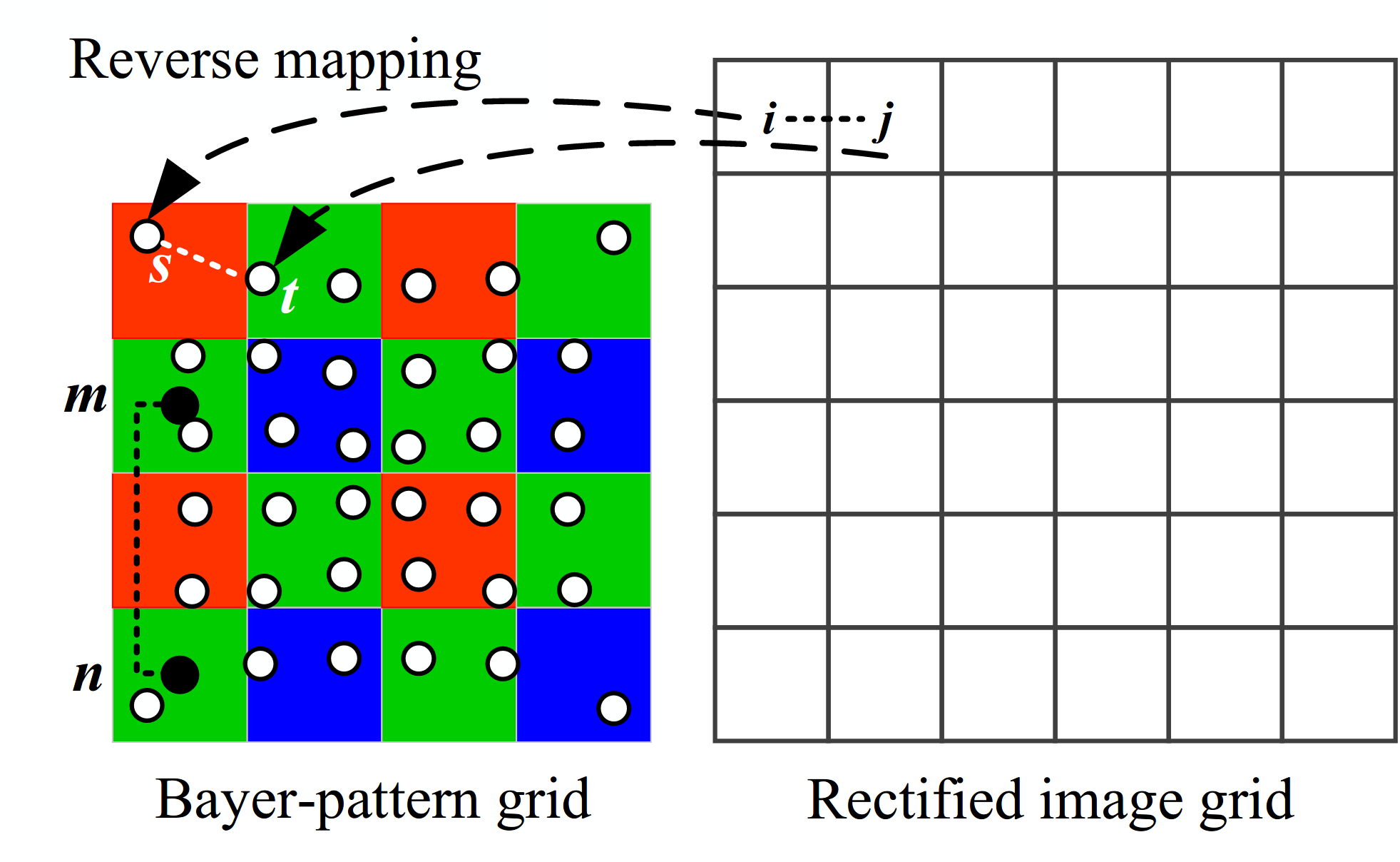}
	\vspace{-0.10in}
	\caption{\small Reverse mapping from the rectified grid to the Bayer-patterned grid. The white circles at the non-integer location are the mapped locations from the rectified grid. The black circles at the integer location represent for captured pixels in the same color channel on the Bayer-patterned grid. Pixel intensities at these locations are used for estimating the weight between the white circles.}
	\label{fig:mapping}
\end{figure}

From OCamCalib \cite{4059340}, an omnidirectional camera calibration toolbox, we obtain a mapping function, $f: \mathbb{Z}^2 \rightarrow \mathbb{R}^2$, that maps a pixel $i$'s integer 2D coordinate $(i_x,i_y)$, $i_x,i_y \in \mathbb{Z}$, in the rectified image grid, to a real 2D coordinate $(s_x,s_y)$, $s_x,s_y \in \mathbb{R}$, in the fisheye camera Bayer-patterned image grid. 
%See Fig.\,\ref{fig:mapping} for an illustration.
See Fig.\,\ref{fig:mapping} for an illustration of reverse pixel mapping from two adjacent pixels $i$ and $j$ in the rectified image grid to locations $s$ and $t$ in the Bayer-patterned grid.

In a conventional demosaicking algorithm \cite{dm}, each color pixel (say red) in an image grid is interpolated as a weighted linear combination of neighboring red pixels.
Similarly, we assume that an $N$-pixel color block in the rectified image grid can be linearly interpolated from an $M$-pixel neighborhood $\y \in \mathbb{R}^M$ in the Bayer-patterned grid as $\H \y$, where $\H \in \mathbb{R}^{N \times M}$ is a weight matrix used for interpolation.
More complex non-linear interpolation methods can also be incorporated into our joint demosaicking / rectification framework---we leave this direction for future work.

% \begin{figure}[htb]
% 	\centering
% 	\includegraphics[width=8cm]{fig/mapping_other.pdf}
% 	\vspace{-0.10in}
% 	\caption{\small Reverse mapping from rectified grid to Bayer-patterned grid.}
% 	\label{fig:mapping}
% \end{figure}

% In general, an interpolated red pixel at integer location $(i,j)$ in the rectified grid is likely to be more accurate if the reverse-mapped non-integer location $(s,t)$ is close to the location of the nearest red captured red pixel $(m,n)$.
% We can thus assign a weight $g_{i,j}$ that quantifies the accuracy of the interpolation as an exponential function of the negative Euclidean distance between $(s,t)$ and $(m,n)$. 
%\red{what function did u use in your experiment?}

\vspace{-0.2in}
\subsection{MAP Formulations for Image Restoration}

To reconstruct a target square pixel patch $\x$ in the rectified grid in a chosen color component, we formulate the following optimization given an $M$-pixel neighborhood $\y$ in the Bayer-patterned grid using GLR \cite{7814302} as follows:

\vspace{-0.1in}
\begin{small}
\begin{align}
\min_{\x} \left \|\H \y - \x \right \|_2^2
% \min_{\x} \left(\H \y - \x \right)^{\top} \G \left(\H \y - \x \right)  
%+ \gamma \; \y^{\top} \L_{\y} \y 
+ \mu \; \x^{\top} \L_{\x} \x 
\label{eq:MAP}
\end{align}
\end{small}\noindent
where $\mu > 0$ is weight parameter to trade off the (first) fidelity term with the (second) signal prior.
% , and $\G$ is a diagonal weight matrix that accounts for the interpolation accuracy $g_{i,j}$ previously discussed. 
In words, \eqref{eq:MAP} states that the reconstructed signal $\x$ should be similar to interpolation $\H \y$ while being smooth with respect to a graph specified by $\L_{\x}$. 

For fixed interpolation matrix $\H$ and graph Laplacian $\L_{\x}$, \eqref{eq:MAP} is an unconstrained quadratic programming (QP) problem with solution computed from a system of linear equations:
\begin{align}
\left( \I + \mu \L_{\x} \right) \x = \H \y
% \left( \G + \mu \L_{\x} \right) \x = \G \H \y
\label{eq:linEq}
\end{align}
Since the coefficient matrix $\I + \mu \, \L_{\x}$ in \eqref{eq:linEq} is sparse, symmetric and positive definite (PD), \eqref{eq:linEq} can be efficiently solved without matrix inverse using a fast numerical linear algebra algorithm such as \textit{conjugate gradient} (CG) \cite{MOLLER1993525}. 

\vspace{-0.05in}
\subsection{Graph Construction}

The restoration performance of \eqref{eq:linEq} depends heavily on how the underlying graph is constructed for target patch $\x$ in the rectified 2D grid, which determines Laplacian $\L_{\x}$. 
For connectivity of patch $\x$, we assume an 8-connected graph, where each pixel in $\x$ is connected to its immediate vertical, horizontal and diagonal neighbors.
For edge weight $w_{i,j}$ that connects pixel pair $(i,j)$ in $\x$, conventionally it is inversely proportional to the \textit{feature distance} of the two corresponding nodes; \textit{i.e.}, the larger the feature distance, the smaller the edge weight \cite{shuman13}. 
In our imaging scenario, we assume that the feature distance is the magnitude of the estimated \textit{signal gradient} $\Delta_{i,j} \in \mathbb{R}$ between samples $i$ and $j$. 
%We define a weighted adjacency matrix $\W$ and each of its entries $w_{i,j}$.
Using an exponential function as the kernel, we can write $w_{i,j}$ as
% \red{do u need magnitude of $\Delta_{ij}$?}
\vspace{-0.1in}
\begin{align}
\label{eq:edge_weight}
w_{i,j} = \exp \left\{- \frac{ \Delta_{i,j}^2 }{\sigma_w^2} \right\}
% w_{i,j} = \exp \left\{- \frac{ \left\| \Delta_{i,j} \right\|^2_2}{\sigma_w^2} \right\}
\end{align}
where $\sigma_w$ is a parameter. 
%We also define a degree matrix $\D$ with $D_{i,i} = \sum_{j=1} u_{i,j}$ and the corresponding combinatorial graph Laplacian matrix is $\L_{\x} = \D - \W$.  
\eqref{eq:edge_weight} implies that $0 \leq w_{i,j} \leq 1$, and the resulting graph Laplacian $\L_{\x}$, as defined in Section\,\ref{sec:prelim}, is positive semi-definite (PSD) (see \cite{8334407} for a proof using Gershgorin Circle Theorem).  

The crux of the graph construction procedure thus rests in the gradient estimation for a pixel pair $(i,j)$.
We estimate gradient $\Delta_{i,j} \in \mathbb{R}$ via a \textit{maximum likelihood estimation} (MLE) formulation as follows. 
Suppose we have access to $K$ noisy observations $\delta_{i,j}^k$ of $\Delta_{i,j}$, $k \in \{1, \ldots, K\}$.
Then MLE of $\Delta_{i,j}$ given $\delta_{i,j}^k$ is:

\vspace{-0.1in}
\begin{small}
\begin{align}
\max_{\Delta_{i,j}} \mathrm{Pr} \left(\Delta_{i,j} ~|~ \{\delta_{i,j}^k\}_{k=1}^K \right) \rightarrow 
\max_{\Delta_{i,j}} \prod_{k=1}^K \mathrm{Pr}
\left( \delta_{i,j}^k ~|~ \Delta_{i,j}
\right) 
\label{eq:MLE}
\end{align}
\end{small}\noindent
where in \eqref{eq:MLE} we assume that each noisy observation $\delta_{i,j}^k$ is generated independently, each with the following distribution:
\begin{align}
\mathrm{Pr} \left( \delta_{i,j}^k | \Delta_{i,j} 
\right) = \exp \left\{ -
v_{i,j}^{k} ( \Delta_{i,j} - \delta_{i,j}^k)^2
% v_{i,j}^{k} \| \Delta_{i,j} - \delta_{i,j}^k \|_2^2
\right\}
\end{align}
where $v_{i,j}^k$ is a unique parameter for random variable $\delta_{i,j}^k$, to be discussed in details.

Minimizing the negative log of likelihood \eqref{eq:MLE}, we get
\vspace{-0.05in}
\begin{align}
\min_{\Delta_{i,j}} &
\sum_{k=1}^{K} v_{i,j}^{k} ( \Delta_{i,j} - \delta_{i,j}^k)^2 
% \sum_{k=1}^{K} v_{i,j}^{k} \left\| \Delta_{i,j} - \delta_{i,j}^k \right\|_2^2
% \nonumber 
% \\
% & = \sum_{k=1}^{K} v_{i,j}^{k} \left( \Delta_{i,j} - \delta_{i,j}^k \right)^{\top}
% \left( \Delta_{i,j} - \delta_{i,j}^k \right)
\label{eq:MLE2}
\end{align}
To solve \eqref{eq:MLE2}, we take the derivative with respect to $\Delta_{i,j}$ and set it to $0$, resulting in
\vspace{-0.1in}
\begin{align}
%\sum_{k=1}^K u_{i,j}^{k} \left( \Delta_{i,j}^* - \delta_{i,j}^k \right) &= \mathbf{0} \nonumber \\
\Delta_{i,j}^* &= \frac{1}{V} \sum_{k=1}^K v_{i,j}^{k} \delta_{i,j}^k
\label{eq:soln}
\end{align}
where $V = \sum_{k=1}^K v_{i,j}^{k}$.
In other words, the solution \eqref{eq:soln} is a weighted average of the noisy gradient observations $\delta_{i,j}^k$.

\vspace{-0.05in}
\subsection{Noise Model for Inter-pixel Gradient}

We obtain $K$ ``noisy" gradient observations $\delta_{i,j}$ in our joint demosaicking / rectification framework as follows. 
We first define a spatial neighborhood  $\mathcal{N}_{i,j}$ surrounding non-integer locations $s$ and $t$ in the Bayer-patterned grid that correspond to pixel pair $(i,j)$ in the rectified grid. 
Within $\mathcal{N}_{i,j}$, we discover $(m,n) \in \mathcal{N}_{i,j}$ that is a pair of adjacent captured pixels of the same color in the Bayer-patterned grid.
% Without loss of generality, assume that the location of pixel $m$ is closer to $s$ than $n$, and the location of pixel $n$ is closer to $t$ than $m$. 
We first compute gradient for pair $(m,n) \in \cN_{i,j}$:
%\vspace{-0.05in}
\begin{align}
\label{eq:bcg}
\delta^{m,n}_{i,j} = y_m - y_n
%\delta_{i,j}^{m,n} = |m_y - n_y|
% \left( s_y - t_y, \theta_{m,n} \right)
\end{align}
where $y_m$ and $y_n$ are the pixel intensity corresponding to the pair $(m,n)$ on the Bayer-pattern $\y$.
% where $\theta_{m,n}$ denotes the angle of the pixel pair $(m,n)$ with respect to the vertical. 

We next compute an associated weight $v_{i,j}^{m,n}$ 
% for $(s,t)\in\mathcal{N}_{i,j}$
as:

\vspace{-0.05in}
\begin{small}
\begin{align}
\label{eq:bcf}
v_{i,j}^{m,n} =
% \sum_{(s,t)\in\mathcal{N}_{i,j}} 
\exp \left\{
- \frac{\| \l_s - \l_m \|_2^2 \;
\| \l_t - \l_n \|_2^2}{\sigma_v^2} \right\}
\; 
\cos \theta^{m,n}_{s,t} \;
\rho_{s,t}^{m,n}
\end{align}
\end{small}\noindent
where $\l_s$ is the coordinate of pixel $s$ in the Bayer-patterned grid. 
% \red{still missing the discussion on why we compare $s$ to $m$ and $t$ to $n$?}
The pairing of $(s, m)$ and $(t, n)$ is determined by a pre-computed lookup table, using the pixel locations of pairs $(m, n)$ and $(s, t)$ as input.
$\theta^{m,n}_{s,t}$ is the angle between line $(s,t)$ and line $(m,n)$, and $\sigma_v$ is a parameter.
$\rho_{s,t}^{m,n}$ is a \textit{color gradient correlation factor} that estimates the correlation of color gradients between colors of pairs $(m,n)$ and $(s,t)$, where pairs $(m,n)$ and $(s,t)$ may belong to different color channels.
% \red{need more explanation of how $\rho$ is computed.}
Specifically, when computing $\rho$ for red and blue channels, we reshape the $M$-pixel Bayer-pattern patch $\y \in \mathbb{R}^{\sqrt{M}\times \sqrt{M}}$ into 4 submatrices $\tilde \y \in \mathbb{R}^{\frac{\sqrt{M}}{2} \times \frac{\sqrt{M}}{2}}$ based on color channels, where within each 4-pixel RGGB array on the Bayer-pattern, pixels of green channel in the diagonal location will be considered as 2 independent channels---Green-1 and Green-2.
In the case of the red channel, the 2-D correlation coefficient between it and the other 3 channel (Blue, Green-1, Green-2) will be computed as $\rho_{RB}, \rho_{RG_1}, \rho_{RG_2}$, and its self-correlation coefficient $\rho_{RR}=1$.
The coefficient of the blue channel will be computed in the same way.
For the green channel, we use the maximum value of the two computed correlation coefficients between it and the other two channels (Red and Blue), \ie, $\rho_{RG} = \max \{\rho_{RG_1}, \rho_{RG_2}\}$ and $\rho_{BG} = \max \{\rho_{BG_1}, \rho_{BG_2}\}$. 
% \red{maybe it's not clearly enough? If so we can draw a figure for the explanation.}
% $\rho$ is calculated is calculated with all captured same color pairs (say all red pairs and green pairs) within the patch.
%Specifically, for each color component (say red), we compute the correlation factor between it and the other two color components, i.e., $\rho_{RG}$ and $\rho_{RB}$. We set \textit{intra-channel} (when color pairs $(i,j)$ and $(s,t)$ are in the same channel) gradient correlation factor $\rho_{RR}$ as 1. %To be updated.
% We heuristically set the \textit{inter-channel} gradient correlation factors so that the red-green and blue-green ones are larger than the red-blue one, since the number of green-channel observations on a Bayer-pattern image patch is twice as that of red- and blue-channel.}
%\red{this is not a complete description of the algorithm. how are the correlation factor $\rho$ computed?}
See Algorithm \ref{alg:proposedalg} for a summary of the proposed algorithm.

\begin{algorithm}[htp]
\caption{Joint demosaicking / rectification method.}
\label{alg:proposedalg}
\textbf{Input}: Bayer-pattern image patch $\y$, $\H$.\\
\textbf{Output}: Target image patch $\x^*$.
\begin{algorithmic}[1]
% \State Initialize $\H\y$ with bilinear method 
\State \textbf{for} \textit{each pair $(i, j)$ on targeted image patch} \textbf{do}
\State $~~~~~$ Locating pair $(s, t)$ on Bayer pattern $\y$ with $\H$
\State $~~~~~$ Compute the correlation factor $\rho$ with observations in $\mathcal{N}_{i,j}$
\State $~~~~~$ Compute the gradient $\delta^{m, n}_{i, j}$ and weight $v_{i, j}^{m, n}$ with each pair $(m, n) \in \mathcal{N}_{i,j}$ in three channels with \eqref{eq:bcg} and \eqref{eq:bcf}.
\State $~~~~~$ For each channel, compute the estimated gradient $\Delta_{i, j}^*$ by weighted average via \eqref{eq:soln}.
\State $~~~~~$ For each $\Delta_{i, j}^*$, compute the edge weight $w_{i, j}$ with \eqref{eq:edge_weight}.
\State \textbf{end for}
\State $~~~~~$ Compute the initial $\L_{\x}$ via $\L_{\x} = \D - \A$.
% \State Initialize $\L_{\x}$ based on $\H\y$, \eqref{eq:edge_weight}, \eqref{eq:soln}, 
% \eqref{eq:bcg}, and \eqref{eq:bcf}. 
\State \textbf{while} \textit{not converge} \textbf{do}
\State $~~~~~$ Solving $\x^*$ via \eqref{eq:linEq}.  
\State $~~~~~$ Update $\L_{\x}^*$ based on $\x^*$ with \eqref{eq:edge_weight}.
\State \textbf{end while}
\end{algorithmic}
\end{algorithm}

\vspace{-0.2in}
\section{Experiments}
\label{sec:results}
\begin{figure*}[!ht]
     \centering
     \begin{subfigure}[b]{0.17\textwidth}
         \centering
             \includegraphics[width=\textwidth]{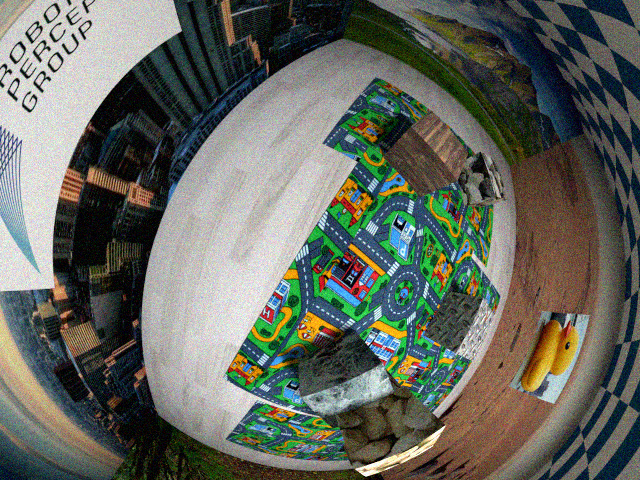}
        %  \caption{}
         \label{fig: fisheye room}
     \end{subfigure}
     \begin{subfigure}[b]{0.17\textwidth}
         \centering
         \includegraphics[width=\textwidth]{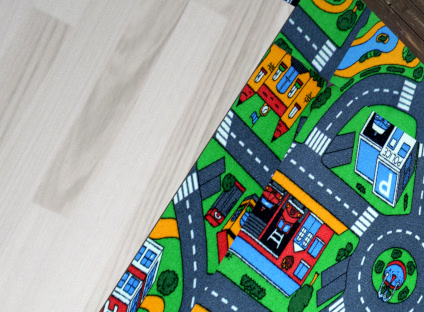}
        %  \caption{}
         \label{fig: gt room}
     \end{subfigure}
     \begin{subfigure}[b]{0.17\textwidth}
         \centering
         \includegraphics[width=\textwidth]{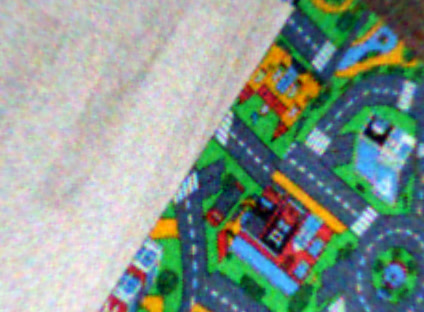}
        %  \caption{}
         \label{fig: bl room}
     \end{subfigure}
     \begin{subfigure}[b]{0.17\textwidth}
         \centering
         \includegraphics[width=\textwidth]{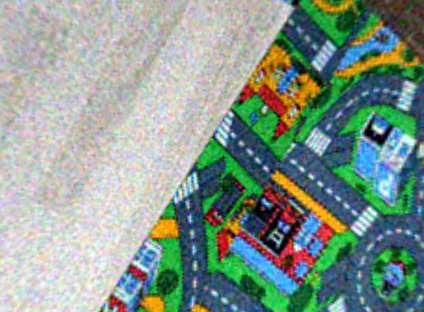}
        %  \caption{}
         \label{fig: mat room}
     \end{subfigure}
     \begin{subfigure}[b]{0.17\textwidth}
         \centering
         \includegraphics[width=\textwidth]{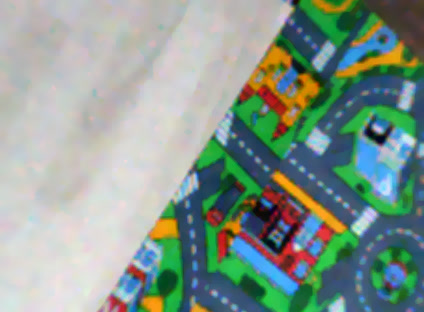}
        %  \caption{}
         \label{fig: proposed room}
     \end{subfigure}
     \\
\vspace{-0.13in}
\ 
\begin{subfigure}[b]{0.17\textwidth}
         \centering
         \includegraphics[width=\textwidth]{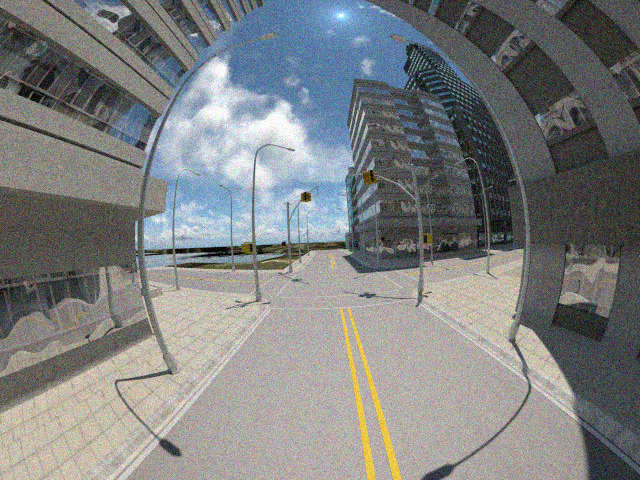}
         \caption{}
         \label{fig: fisheye city}
     \end{subfigure}
     \begin{subfigure}[b]{0.17\textwidth}
         \centering
         \includegraphics[width=\textwidth]{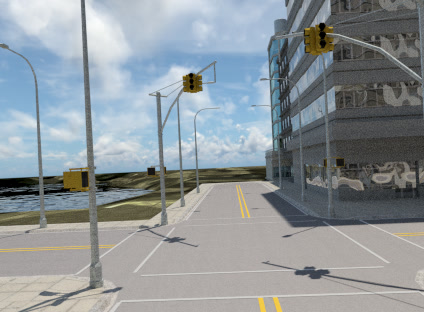}
         \caption{}
         \label{fig: gt city}
     \end{subfigure}
     \begin{subfigure}[b]{0.17\textwidth}
         \centering
         \includegraphics[width=\textwidth]{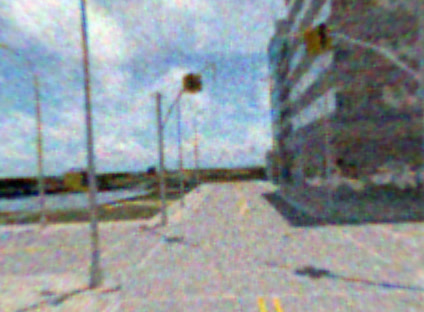}
         \caption{}
         \label{fig: bl city}
     \end{subfigure}
     \begin{subfigure}[b]{0.17\textwidth}
         \centering
         \includegraphics[width=\textwidth]{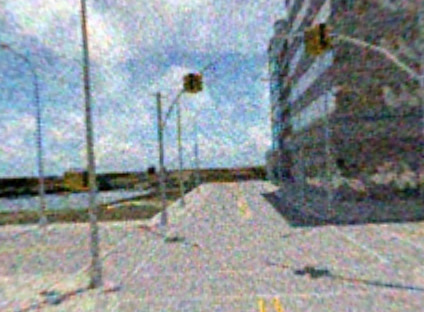}
         \caption{}
         \label{fig: mat city}
     \end{subfigure}
     \begin{subfigure}[b]{0.17\textwidth}
         \centering
         \includegraphics[width=\textwidth]{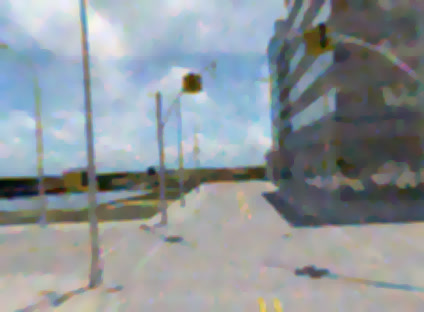}
         \caption{}
         \label{fig:Muti-FoV proposed}
     \end{subfigure}
\vspace{-0.1in}     
\caption{Results of demosaicking and rectification for \texttt{room} and \texttt{city} \cite{7487210}. (a) Ground truth fisheye camera image. (b) Ground truth pinhole image. (c) Demosaicking and rectification using the bilinear method. (d) Demosaicking using high quality linear interpolation (HQL) \cite{1326587} and rectification using the bilinear method. (e) Our proposed joint demosaicking / rectification method.}
\label{fig:Multi-Fov}
\end{figure*}

% \begin{figure*}[!ht]
%      \centering
%      \begin{subfigure}[b]{0.19\textwidth}
%          \centering
%          \includegraphics[width=\textwidth]{fig/city_input.jpg}
%          \caption{}
%          \label{fig: fisheye city}
%      \end{subfigure}
%      \hfill
%      \begin{subfigure}[b]{0.19\textwidth}
%          \centering
%          \includegraphics[width=\textwidth]{fig/city_gt.jpg}
%          \caption{}
%          \label{fig: gt city}
%      \end{subfigure}
%      \hfill
%      \begin{subfigure}[b]{0.19\textwidth}
%          \centering
%          \includegraphics[width=\textwidth]{fig/city_bl.jpg}
%          \caption{}
%          \label{fig: bl city}
%      \end{subfigure}
%      \hfill
%      \begin{subfigure}[b]{0.19\textwidth}
%          \centering
%          \includegraphics[width=\textwidth]{fig/city_mat.jpg}
%          \caption{}
%          \label{fig: mat city}
%      \end{subfigure}
%      \hfill
%      \begin{subfigure}[b]{0.19\textwidth}
%          \centering
%          \includegraphics[width=\textwidth]{fig/city_joint.jpg}
%          \caption{}
%          \label{fig: proposed city}
%      \end{subfigure}
% \vspace{-0.1in}
% \caption{Results of demosaicking and rectification for \texttt{city}. (a) Ground truth fisheye camera image. (b) Ground truth pinhole image. (c) Demosaicking and rectification using the bilinear method. (d) Demosaicking using HQL interpolation \cite{1326587} and rectification using the bilinear method. (e) Our proposed joint demosaicking / rectification method.}
%         \label{fig:scene room}
% \end{figure*}

\begin{figure*}[!htb]
     \centering
     \begin{subfigure}[b]{0.15\textwidth}
         \centering
         \includegraphics[width=\textwidth]{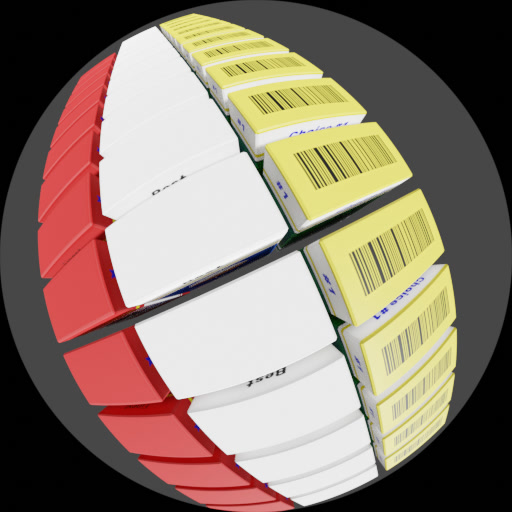}
         \label{fig:F1}
     \end{subfigure}
     \begin{subfigure}[b]{0.15\textwidth}
         \centering
         \includegraphics[width=\textwidth]{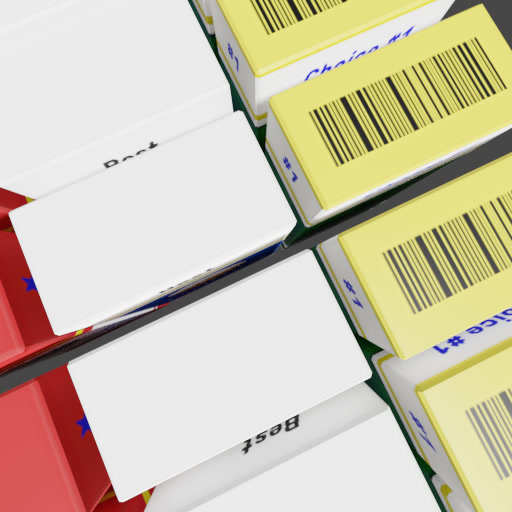}
         \label{fig:G1}
     \end{subfigure}
     \begin{subfigure}[b]{0.15\textwidth}
         \centering
         \includegraphics[width=\textwidth]{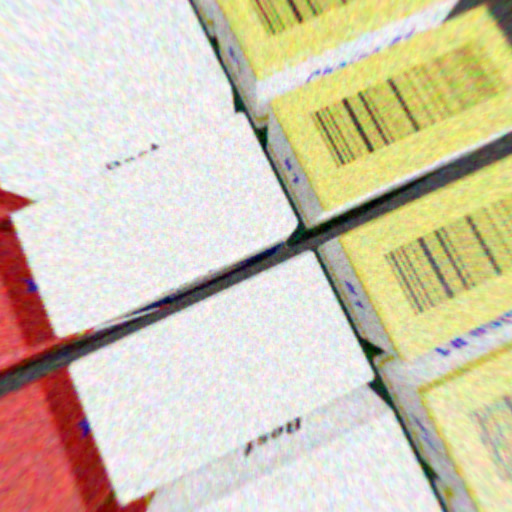}
         \label{fig:B1}
     \end{subfigure}
     \begin{subfigure}[b]{0.15\textwidth}
         \centering
         \includegraphics[width=\textwidth]{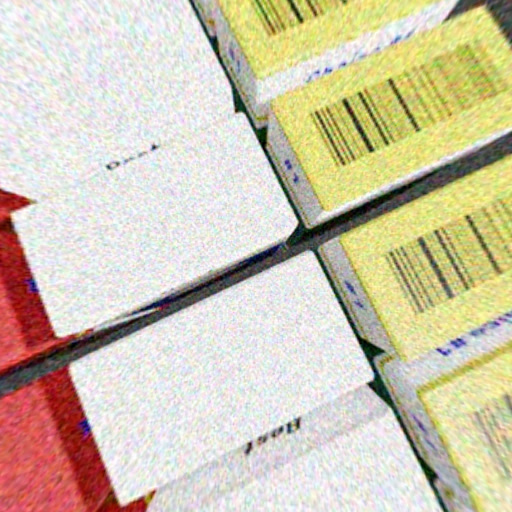}
         \label{fig:M1}
     \end{subfigure}
     \begin{subfigure}[b]{0.15\textwidth}
         \centering
         \includegraphics[width=\textwidth]{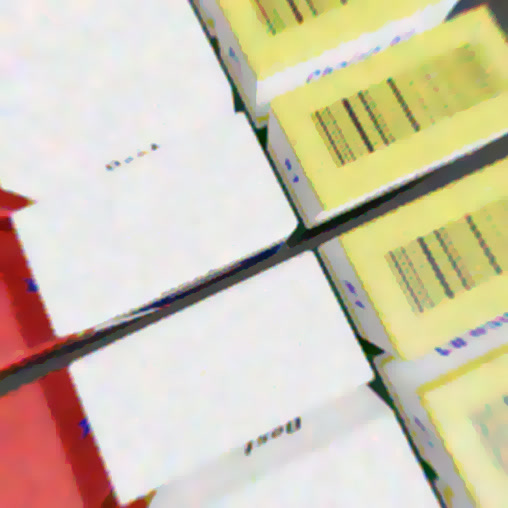}
         \label{fig:P1}
     \end{subfigure}\\
     \vspace{-0.13in}
     \begin{subfigure}[b]{0.15\textwidth}
         \centering
         \includegraphics[width=\textwidth]{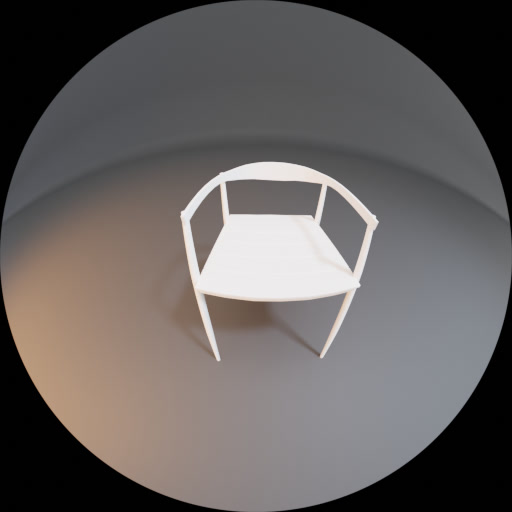}
         \label{fig:F2}
     \end{subfigure}
     \begin{subfigure}[b]{0.15\textwidth}
         \centering
         \includegraphics[width=\textwidth]{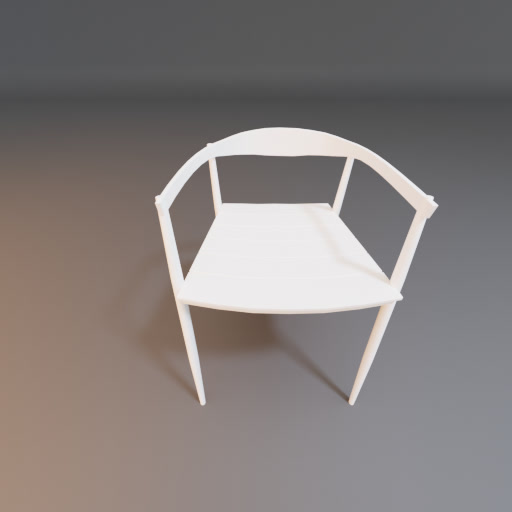}
         \label{fig:G2}
     \end{subfigure}
     \begin{subfigure}[b]{0.15\textwidth}
         \centering
         \includegraphics[width=\textwidth]{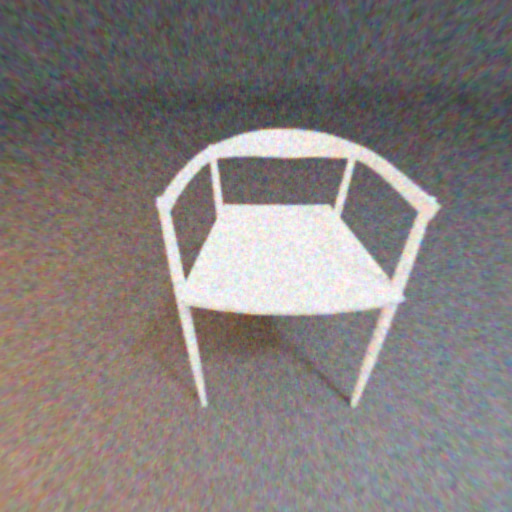}
         \label{fig:B2}
     \end{subfigure}
     \begin{subfigure}[b]{0.15\textwidth}
         \centering
         \includegraphics[width=\textwidth]{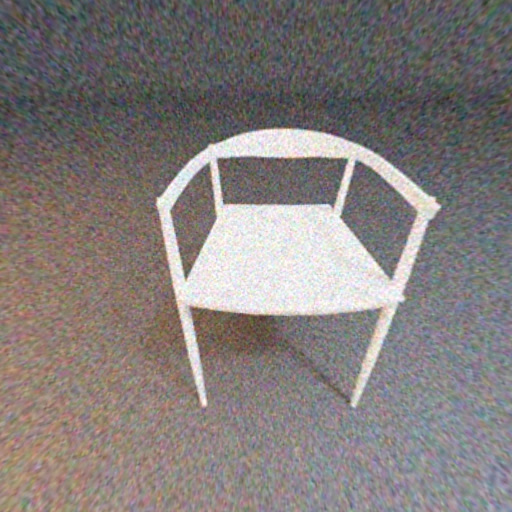}
         \label{fig:M2}
     \end{subfigure}
     \begin{subfigure}[b]{0.15\textwidth}
         \centering
         \includegraphics[width=\textwidth]{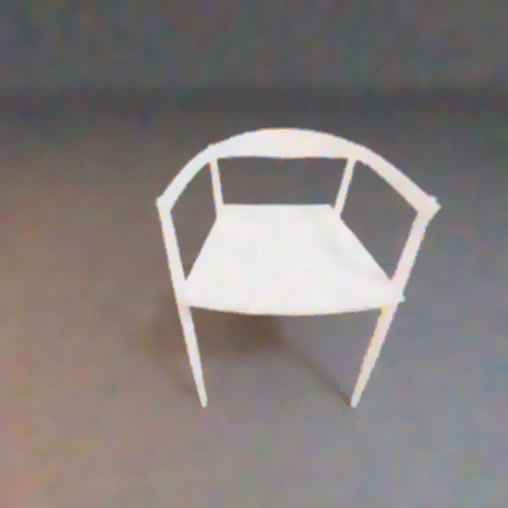}
         \label{fig:P2}
     \end{subfigure}\\
      \vspace{-0.13in}
     \begin{subfigure}[b]{0.15\textwidth}
         \centering
         \includegraphics[width=\textwidth]{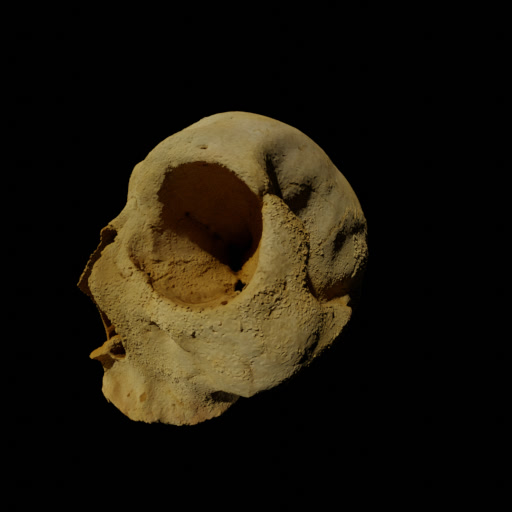}
         \label{fig:F3}
     \end{subfigure}
     \begin{subfigure}[b]{0.15\textwidth}
         \centering
         \includegraphics[width=\textwidth]{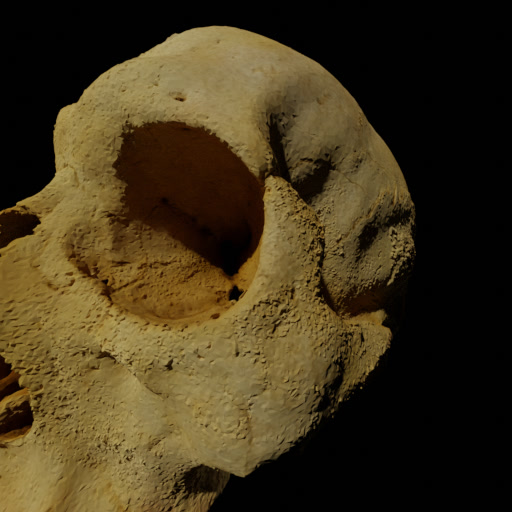}
         \label{fig:G3}
     \end{subfigure}
     \begin{subfigure}[b]{0.15\textwidth}
         \centering
         \includegraphics[width=\textwidth]{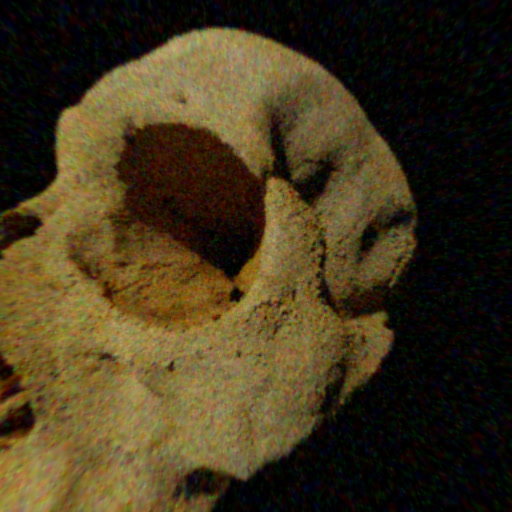}
         \label{fig:B3}
     \end{subfigure}
     \begin{subfigure}[b]{0.15\textwidth}
         \centering
         \includegraphics[width=\textwidth]{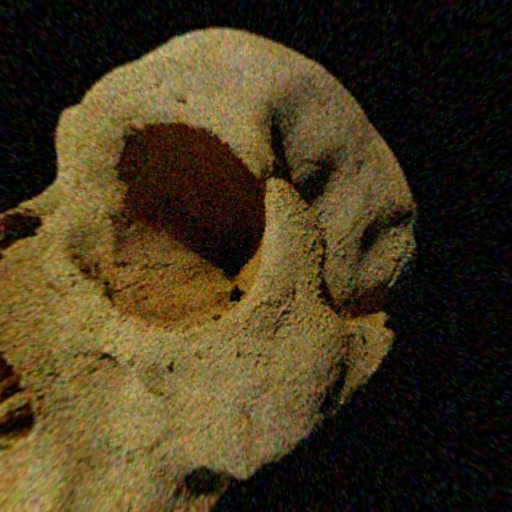}
         \label{fig:M3}
     \end{subfigure}
     \begin{subfigure}[b]{0.15\textwidth}
         \centering
         \includegraphics[width=\textwidth]{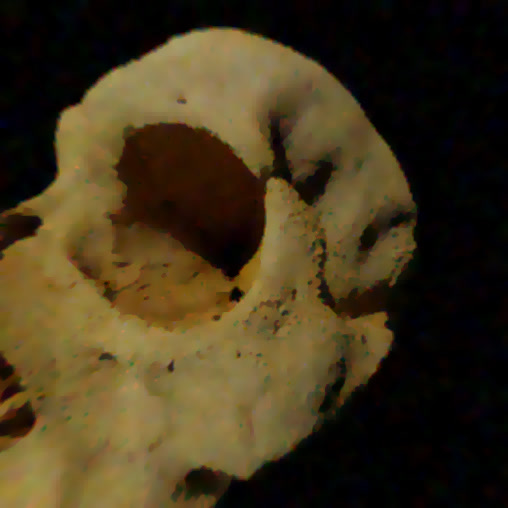}
         \label{fig:P3}
     \end{subfigure}
     \\
          \vspace{-0.13in}
          \ 
     \begin{subfigure}[b]{0.15\textwidth}
         \centering
     \includegraphics[width=\textwidth]{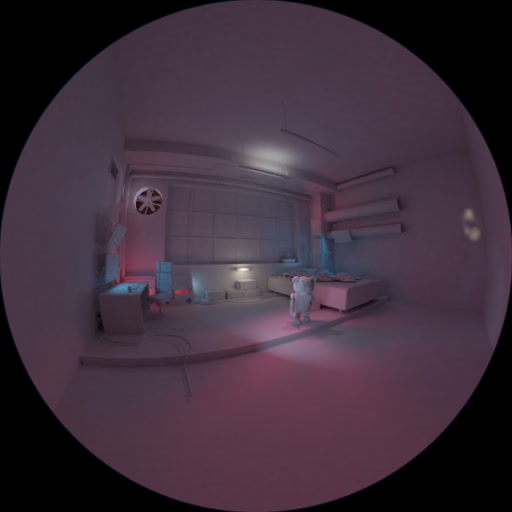}
         \caption{}
         \label{fig:F4}
     \end{subfigure}
     \begin{subfigure}[b]{0.15\textwidth}
         \centering
         \includegraphics[width=\textwidth]{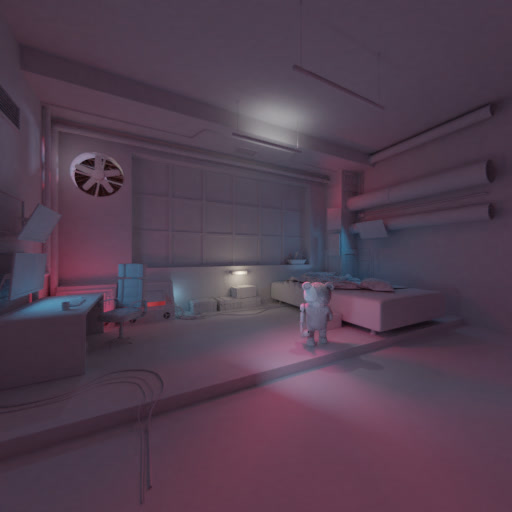}
         \caption{}
         \label{fig:G4}
     \end{subfigure}
     \begin{subfigure}[b]{0.15\textwidth}
         \centering
         \includegraphics[width=\textwidth]{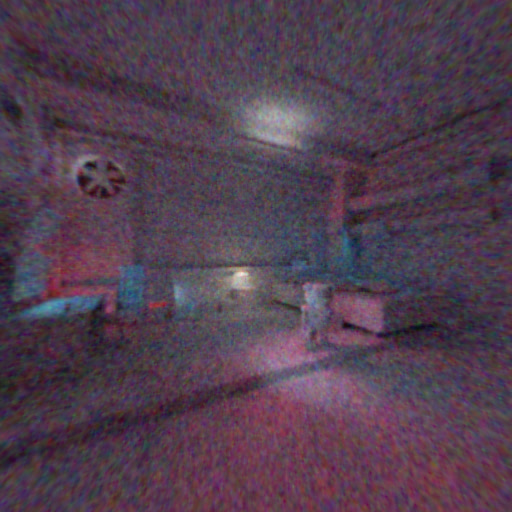}
         \caption{}
         \label{fig:B4}
     \end{subfigure}
     \begin{subfigure}[b]{0.15\textwidth}
         \centering
         \includegraphics[width=\textwidth]{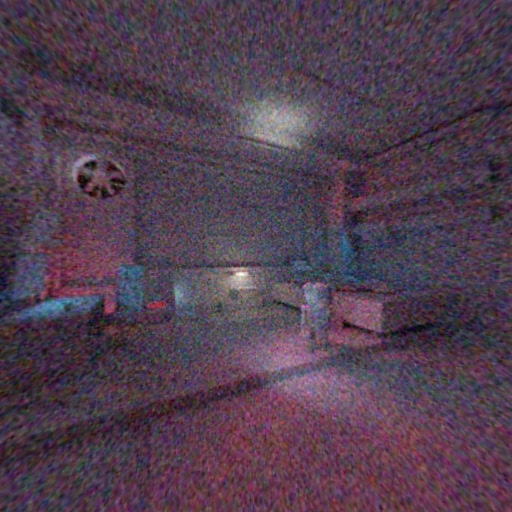}
         \caption{}
         \label{fig:M4}
     \end{subfigure}
     \begin{subfigure}[b]{0.15\textwidth}
         \centering
         \includegraphics[width=\textwidth]{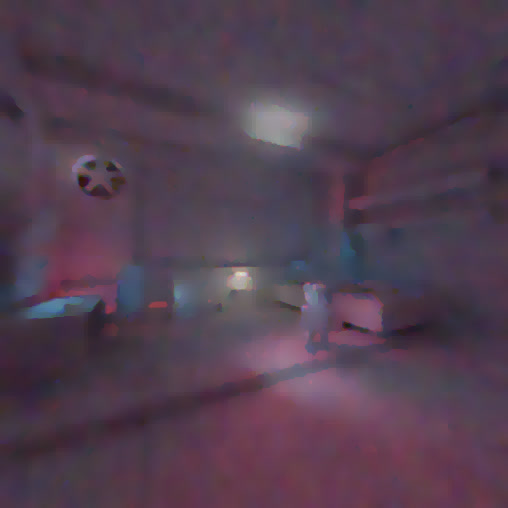}
         \caption{}
         \label{fig:P4}
     \end{subfigure}
\vspace{-0.1in}
\caption{Demosaicking and rectification result of the in-house dataset, where the images are generated  from the 3-D models: \texttt{box}, \texttt{chair}, \texttt{skull} and \texttt{teddy}. (a) Ground truth fisheye camera image. (b) Ground truth pinhole image. (c) Demosaicking and rectification using the bilinear method. (d) Demosaicking using HQL interpolation \cite{1326587} and rectification using the bilinear method. (e) Our proposed joint demosaicking / rectification method.}
        \label{fig:in-house}
\end{figure*}

\subsection{Experimental Setup}
%\vspace{-0.05in}
We tested our joint demosaicking / rectification algorithm on a Multi-FoV image dataset \cite{7487210} and our in-house constructed dataset\footnote{The dataset is available at: 
% \url{https://github.com/fengbolan/GISP-Fisheye-Rectification-Dataset}}
\href{https://github.com/fengbolan/York-Fisheye-Image-Rectification-Dataset}{https://github.com/fengbolan/York-Fisheye-Image-Rectification-Dataset} }. 
The Multi-FoV image dataset includes two scenes: \texttt{room} and \texttt{city}. 
5 images from \texttt{room} and 25 images from \texttt{city} were used in our experiment.
Our in-house dataset includes $140$ pinhole and fisheye camera images generated from 4 publicly available 3-D models: \texttt{box}, \texttt{chair}, \texttt{skull} and \texttt{teddy}.
3 images from each scene were used for evaluation.
For demosaicking, we employed two competing schemes: 1) bilinear interpolation, and 2) a high quality linear (HQL) filter \cite{1326587}. 
For rectification, we employed a bilinear interpolation method. 
Given a fisheye image, as depicted in Fig.\,\ref{fig: fisheye city} and \ref{fig:F4}, we designated a image region that corresponded to the ground truth image (rectified image) as the region of interest (ROI), shown in Fig.\,\ref{fig: gt city} and \ref{fig:G4}, respectively.
We then removed color pixels in the ROI to generate a Bayer-patterned image with additive Gaussian noise (variance $\sigma = 15$) as the input of competing algorithms.

The parameters of our algorithm were set empirically according to the content of the images. 
$\mu$ was set to $1$ in all settings. 
We used $5$ iterations for the Multi-FoV dataset images and $8$ iterations for the proposed  dataset images.
$\sigma_w$ was set to $0.01$ in the first iteration and $0.02$ in the remaining iterations for the images from the Multi-FoV dataset, and it was set to $0.035$ in the first iteration and $0.028$ in the remaining iterations for our in-house dataset. 
$\sigma_u$ was set to $1.5$ for the Multi-FoV dataset and $6$ for our in-house dataset. 
The patch size was set to $32$ pixels with a stride of $28$ pixels.
The experiments were conducted with Matlab R2019a and a computer with a CPU of intel i7-9700T and 32G of RAM.

\vspace{-0.1in}
\subsection{Quantitative Comparisons}
% \begin{table}[htbp]
%   \centering
%   \caption{Demosaicking and rectification performance of 5 images from scene \texttt{room} under noise level $\sigma = 15$.}
%   \vspace{-0.1in}
%   \begin{adjustbox}{max width=8cm}
%     \begin{tabular}{cccc}
%     \toprule
%           & Bilinear & High Quality Linear \cite{1326587} & Proposed \\
%     \midrule
%     \multicolumn{1}{c}{PSNR} & 20.763 & \textbf{21.041} & 20.910 \\
%     \multicolumn{1}{c}{SSIM\cite{wang04}} & 0.710 & 0.702 & \textbf{0.788} \\
%     \bottomrule
%     \end{tabular}%
%     \end{adjustbox}
%   \label{tab:psnr ssim room}%
% \end{table}%

% \begin{table}[htbp]
%   \centering
%   \caption{Demosaicking and rectification performance of 25 images from scene \texttt{city} under noise level $\sigma = 15$.}
%   \vspace{-0.1in}
%     \begin{adjustbox}{max width=8cm}

%     \begin{tabular}{cccc}
%     \toprule
%           & Bilinear & High Quality Linear \cite{1326587}  & Proposed \\
%     \midrule
%     \multicolumn{1}{c}{PSNR} & 24.237 & 24.245 & \textbf{24.766} \\
%     \multicolumn{1}{c}{SSIM\cite{wang04}} & 0.550 & 0.557 & \textbf{0.622} \\
%     \bottomrule
%     \end{tabular}%
%         \end{adjustbox}

%   \label{tab:psnr ssim city}%
% \end{table}%

%\begin{table}[!ht]
\begin{table}
\centering
\caption{
% Demosaicking and rectification performance 
Average SSIM \cite{wang04} and PSNR of images from 6 scenes under noise level $\sigma = 15$. 5 and 25 images from \texttt{room} and \texttt{city} respectively from the Multi-FoV image dataset \cite{7487210} were used. 
3 images each from scene \texttt{box}, \texttt{chair}, \texttt{skull} and \texttt{teddy} from our in-house dataset were used. 
For demosaicking, bilinear interpolation and high quality linear (HQL) method \cite{1326587} were used for comparison. 
In both cases, bilinear was used for rectification.}
\resizebox{\columnwidth}{!}{
\begin{tabular}{ccccccc}
\hline
\multicolumn{1}{|c|}{\multirow{2}{*}{Scene name}} & \multicolumn{3}{c|}{SSIM \cite{wang04}}& \multicolumn{3}{c|}{PSNR (dB)}\\ \cline{2-7} 
\multicolumn{1}{|c|}{}                            & \multicolumn{1}{c|}{Bilinear} & \multicolumn{1}{c|}{HQL\cite{1326587}} & \multicolumn{1}{c|}{Proposed}       & \multicolumn{1}{c|}{Bilinear} & \multicolumn{1}{c|}{HQL\cite{1326587}} & \multicolumn{1}{c|}{Proposed}        \\ \hline
\hline
\multicolumn{1}{|c|}{\texttt{room} \cite{7487210}}                        & \multicolumn{1}{c|}{0.710}    & \multicolumn{1}{c|}{0.702}                 & \multicolumn{1}{c|}{\textbf{0.788}} & \multicolumn{1}{c|}{20.76}   & \multicolumn{1}{c|}{\textbf{21.04}}       & \multicolumn{1}{c|}{20.91}          \\ \hline
\multicolumn{1}{|c|}{\texttt{city} \cite{7487210}}                        & \multicolumn{1}{c|}{0.550}    & \multicolumn{1}{c|}{0.557}                 & \multicolumn{1}{c|}{\textbf{0.622}} & \multicolumn{1}{c|}{24.24}   & \multicolumn{1}{c|}{24.25}                & \multicolumn{1}{c|}{\textbf{24.77}} \\ \hline
\multicolumn{1}{|c|}{\texttt{box}}                         & \multicolumn{1}{c|}{0.599}    & \multicolumn{1}{c|}{0.531}                 & \multicolumn{1}{c|}{\textbf{0.849}} & \multicolumn{1}{c|}{21.88}   & \multicolumn{1}{c|}{21.20}                & \multicolumn{1}{c|}{\textbf{22.52}} \\ \hline
\multicolumn{1}{|c|}{\texttt{chair}}                       & \multicolumn{1}{c|}{0.601}    & \multicolumn{1}{c|}{0.505}                 & \multicolumn{1}{c|}{\textbf{0.916}} & \multicolumn{1}{c|}{26.68}   & \multicolumn{1}{c|}{25.35}                & \multicolumn{1}{c|}{\textbf{29.80}} \\ \hline
\multicolumn{1}{|c|}{\texttt{skull}}                       & \multicolumn{1}{c|}{0.648}    & \multicolumn{1}{c|}{0.556}                 & \multicolumn{1}{c|}{\textbf{0.861}} & \multicolumn{1}{c|}{26.02}   & \multicolumn{1}{c|}{24.92}                & \multicolumn{1}{c|}{\textbf{27.58}} \\ \hline
\multicolumn{1}{|c|}{\texttt{teddy}}                       & \multicolumn{1}{c|}{0.722}    & \multicolumn{1}{c|}{0.641}                 & \multicolumn{1}{c|}{\textbf{0.919}} & \multicolumn{1}{c|}{27.63}   & \multicolumn{1}{c|}{26.10}                & \multicolumn{1}{c|}{\textbf{31.63}} \\ \hline
\end{tabular}
}
\label{tab: average PSNR and SSIM}
\end{table}

% \begin{figure}
%      \centering
%      \begin{subfigure}[b]{0.23\textwidth}
%          \centering
%          \includegraphics[width=\textwidth]{fig/room_bayer.jpg}
%          \caption{}
%          \label{fig:}
%      \end{subfigure}
%      \hfill
%      \begin{subfigure}[b]{0.23\textwidth}
%          \centering
%          \includegraphics[width=\textwidth]{fig/city_bayer.jpg}
%          \caption{}
%          \label{fig:}
%      \end{subfigure}
% \vspace{-0.1in}
% \caption{Bayer-patterned images are generated by downsampling the fisheye image of Multi-FoV dataset \cite{7487210}, which is fed to the algorithm for demosaicking and rectification.
% % \red{why would u show the Bayer-patterned but rectified images? if u want to show the Bayer-patterned images, they should be from the fisheye camera perspective, e.g., Fig.3(a) and Fig.4(a). this is super-confusing.}
% }
% \label{fig:bayer}
% \end{figure}

The visual results for \texttt{room} and \texttt{city} are shown in Fig.\,\ref{fig:Multi-Fov} and the results of the in-house dataset are shown in Fig.\,\ref{fig:in-house} respectively. 
The numerical results in average SSIM \cite{wang04} and PSNR are shown in Table.\,\ref{tab: average PSNR and SSIM}.
In Fig.\,\ref{fig:Muti-FoV proposed}, we observe that due to the proposed smoothness prior employed, compared with the other two methods, the result using our proposed method appears smoother while the boundaries in the image were well preserved.
Similar results can be also observed in Fig.\,\ref{fig:B4}-\ref{fig:P4}, where the noise was noticeable on the image with bilinear and HQL methods, but our proposed method was less affected by such noise.
Note that noise in the image demosaicked using the HQL method was more noticeable than the one using the bilinear method. 
This is because the HQL method employs filters to calculate a gradient map to enhance edges in the image.
However, it may also lead to noise enhancement and image quality deterioration.
In contrast, our proposed algorithm achieves a good trade-off between edge enhancement and noise reduction. 

Such a conclusion is supported by Table.\,\ref{tab: average PSNR and SSIM}. 
Although PSNR of the HQL method outperformed the bilinear method on the Multi-FoV dataset images, its SSIM was worse than the other two methods. 
For the Multi-FoV dataset, PSNR of our proposed method for the \texttt{room} images was better than the bilinear method, and it was close to the PSNR of the HQL method. 
SSIM of our proposed method was higher than the other two methods. For the \texttt{city} images, PSNR and SSIM of our proposed method were better than the other two methods.
The proposed method outperformed the other two methods by up to $0.52$ dB in PSNR and $0.086$ in SSIM, respectively.
For our in-house dataset, the proposed method performed well across all images. 
It outperformed the other two competing methods by up to $5.53$dB in PSNR on the images from the scene \texttt{teddy}, and up to $0.411$ in SSIM on the images from scene \texttt{box}.

%\vspace{-0.1in}
%\section{Conclusion}
%\label{sec:conclude}
%\input{conclude.tex}

\bibliographystyle{IEEEbib}
\bibliography{ref,ref2}

\begin{thebibliography}{10}

\bibitem{8337839}
J.~Tan, G.~Cheung, and R.~Ma,
\newblock ``360-degree virtual-reality cameras for the masses,''
\newblock {\em IEEE MultiMedia}, vol. 25, no. 1, pp. 87--94, Jan. 2018.

\bibitem{784434}
R.~{Kimmel},
\newblock ``Demosaicing: image reconstruction from color ccd samples,''
\newblock {\em IEEE Transactions on Image Processing}, vol. 8, no. 9, pp.
  1221--1228, Sep. 1999.

\bibitem{1658081}
K.~{Hirakawa} and T.~W. {Parks},
\newblock ``Joint demosaicing and denoising,''
\newblock {\em IEEE Transactions on Image Processing}, vol. 15, no. 8, pp.
  2146--2157, Aug 2006.

\bibitem{4287011}
L.~{Zhang}, X.~{Wu}, and D.~{Zhang},
\newblock ``Color reproduction from noisy cfa data of single sensor digital
  cameras,''
\newblock {\em IEEE Transactions on Image Processing}, vol. 16, no. 9, pp.
  2184--2197, Sep. 2007.

\bibitem{Gharbi:2016:DJD:2980179.2982399}
M.~Gharbi, G.~Chaurasia, S.~Paris, and F.~Durand,
\newblock ``Deep joint demosaicking and denoising,''
\newblock {\em ACM Trans. Graph.}, vol. 35, no. 6, pp. 191:1--191:12, Nov.
  2016.

\bibitem{DBLP:journals/corr/abs-1802-04723}
W.~Dong, M.~Yuan, X.~Li, and G.~Shi,
\newblock ``Joint demosaicing and denoising with perceptual optimization on a
  generative adversarial network,''
\newblock {\em CoRR}, vol. abs/1802.04723, 2018.

\bibitem{7814302}
J.~Pang and G.~Cheung,
\newblock ``Graph {Laplacian} regularization for image denoising: Analysis in
  the continuous domain,''
\newblock {\em IEEE Transactions on Image Processing}, vol. 26, no. 4, pp.
  1770--1785, April 2017.

\bibitem{7487210}
{Z. Zhang}, H.~{Rebecq}, C.~{Forster}, and D.~{Scaramuzza},
\newblock ``Benefit of large field-of-view cameras for visual odometry,''
\newblock in {\em 2016 IEEE International Conference on Robotics and Automation
  (ICRA)}, May 2016, pp. 801--808.

\bibitem{7740964}
X.~{Liu}, G.~{Cheung}, X.~{Wu}, and D.~{Zhao},
\newblock ``Random walk graph {Laplacian}-based smoothness prior for soft
  decoding of {JPEG} images,''
\newblock {\em IEEE Transactions on Image Processing}, vol. 26, no. 2, pp.
  509--524, Feb. 2017.

\bibitem{hu2015graph}
W.~Hu, G.~Cheung, and M.~Kazui,
\newblock ``Graph-based dequantization of block-compressed piecewise smooth
  images,''
\newblock {\em IEEE Signal Processing Letters}, vol. 23, no. 2, pp. 242--246,
  2015.

\bibitem{wan2016image}
P.~Wan, G.~Cheung, D.~Florencio, C.~Zhang, and O.~C Au,
\newblock ``Image bit-depth enhancement via maximum a posteriori estimation of
  ac signal,''
\newblock {\em IEEE Transactions on Image Processing}, vol. 25, no. 6, pp.
  2896--2909, 2016.

\bibitem{8488519}
Y.~{Bai}, G.~{Cheung}, X.~{Liu}, and W.~{Gao},
\newblock ``Graph-based blind image deblurring from a single photograph,''
\newblock {\em IEEE Transactions on Image Processing}, vol. 28, no. 3, pp.
  1404--1418, March 2019.

\bibitem{8476571}
X.~{Liu}, G.~{Cheung}, X.~{Ji}, D.~{Zhao}, and W.~{Gao},
\newblock ``Graph-based joint dequantization and contrast enhancement of poorly
  lit {JPEG} images,''
\newblock {\em IEEE Transactions on Image Processing}, vol. 28, no. 3, pp.
  1205--1219, March 2019.

\bibitem{ortega18ieee}
A.~Ortega, P.~Frossard, J.~Kovacevic, J.~M.~F. Moura, and P.~Vandergheynst,
\newblock ``Graph signal processing: Overview, challenges, and applications,''
\newblock in {\em Proceedings of the {IEEE}}, May 2018, vol. 106, no.5, pp.
  808--828.

\bibitem{4059340}
D.~{Scaramuzza}, A.~{Martinelli}, and R.~{Siegwart},
\newblock ``A toolbox for easily calibrating omnidirectional cameras,''
\newblock in {\em 2006 IEEE/RSJ International Conference on Intelligent Robots
  and Systems}, Oct 2006, pp. 5695--5701.

\bibitem{dm}
A.~Davies and P.~Fennessy,
\newblock {\em Digital imaging for photographers},
\newblock Focal Press, 2001.

\bibitem{MOLLER1993525}
M.~F. Møller,
\newblock ``A scaled conjugate gradient algorithm for fast supervised
  learning,''
\newblock {\em Neural Networks}, vol. 6, no. 4, pp. 525 -- 533, 1993.

\bibitem{shuman13}
D.~I. Shuman~et al.,
\newblock ``The emerging field of signal processing on graphs: Extending
  high-dimensional data analysis to networks and other irregular domains,''
\newblock in {\em IEEE Signal Processing Magazine}, May 2013, vol.~30, pp.
  83--98.

\bibitem{8334407}
G.~{Cheung}, E.~{Magli}, Y.~{Tanaka}, and M.~K. {Ng},
\newblock ``Graph spectral image processing,''
\newblock {\em Proceedings of the IEEE}, vol. 106, no. 5, pp. 907--930, May
  2018.

\bibitem{1326587}
H.~S. {Malvar}, {L. He}, and R.~{Cutler},
\newblock ``High-quality linear interpolation for demosaicing of
  bayer-patterned color images,''
\newblock in {\em 2004 IEEE International Conference on Acoustics, Speech, and
  Signal Processing}, May 2004, vol.~3, pp. iii--485.

\bibitem{wang04}
Z.~Wang, A.~Bovik, H.~Sheikh, and E.~Simoncelli,
\newblock ``Image quality assessment: From error visibility to structural
  similarity,''
\newblock in {\em IEEE Transactions on Image Processing}, August 2005, vol. 13,
  no.4, pp. 600--612.

\end{thebibliography}

\end{document}